\documentclass[preprint,showpacs,preprintnumbers,amsmath,amssymb]{revtex4}

\usepackage{graphicx}
\usepackage{dcolumn}
\usepackage{bm}
\usepackage{epsf}

\renewcommand{\bar}[1]{\overline{#1}}

\renewcommand{\d}{{\mathrm d}}

\providecommand{\Journal}[4] {#1 {\bf #2} (#4) #3}
\providecommand{\EPJC}{Eur. Phys. J. C } %
\providecommand{\NPB}{Nucl. Phys. B } %
\providecommand{\PL}{Phys. Lett. } %
\providecommand{\PLB}{Phys. Lett. B } %
\providecommand{\PRL}{Phys. Rev. Lett. } %
\providecommand{\PRC}{Phys. Rev. C } %
\providecommand{\PRD}{Phys. Rev. D } %
\providecommand{\RMP}{Rev. Mod. Phys. } %
\providecommand{\RMP}{Rev. Mod. Phys. } %
\providecommand{\ZPC}{Z. Phys. C } %

\begin{document}

\begin{flushright}
Preprint USM-TH-142, to appear in Phys. Lett. B
\end{flushright}

\title{Particle-Antiparticle Asymmetries of
$\Lambda$ Production  in Hadron-Nucleon Collisions}

\author{Bo-Qiang Ma}
\affiliation{Department of Physics, Peking University, Beijing
100871, China}
\author{Ivan Schmidt}
\affiliation{Departamento de F\'\i sica, Universidad T\'ecnica
Federico Santa Mar\'\i a, Casilla 110-V, 
Valpara\'\i so, Chile}
\author{Jian-Jun Yang}
\affiliation{Departamento de F\'\i sica, Universidad T\'ecnica
Federico Santa Mar\'\i a, Casilla 110-V, 
Valpara\'\i so, Chile} \affiliation{Institut f\"ur Theoretische
Physik, Universit\"at Regensburg, D-93040 Regensburg, Germany}
\affiliation{Department of Physics, Nanjing Normal University,
Nanjing 210097, China}

\begin{abstract}
The particle-antiparticle asymmetries of $\Lambda$ production in
250~GeV/c $\pi^{\pm}$, $K^{\pm}$, and $p$ --nucleon collisions are
studied with two model parametrizations of quark to $\Lambda$
fragmentation functions. It is shown that the available data can
be qualitatively explained by the calculated results in both the
quark-diquark model and a pQCD based analysis of fragmentation
functions. The differences in the two model predictions are
significant for $K^{\pm}$ beams, and high precision measurements
of the asymmetries with detailed $x_F$ and $P_T$ information can
discriminate between different predictions.
\end{abstract}


\vfill

\vfill


\vfill

\vfill 

\maketitle


Much of the recent interest in the structure of the $\Lambda$
hyperon stems from the fact that its quark structure is rather
simple in the naive constituent quark model, since it contains
three different quark flavors ($u$, $d$, $s$), and furthermore its
spin is carried only by the strange quark. Therefore any departure
from this picture signals the presence of relativistic and
non-perturbative effects. Experimentally the most effective means
to investigate the $\Lambda$ structure is through the quark to
$\Lambda$ fragmentation in various processes
\cite{MSSY01,MSY00PLB,MSY00Z,ALEPH,OPAL,HERMES,E665,NOMAD,E769}.
Unfortunately there are still no available measurements about the
relations between different flavor-dependent quark to $\Lambda$
fragmentation functions, such as the relation between the favored
and unfavored fragmentation functions, and the relation between
the flavor structure of the favored fragmentation functions
$D_u^{\Lambda}(z)$ and $D_s^{\Lambda}(z)$. Nevertheless there has
been recent new measurement of the $\Lambda$-$\bar{\Lambda}$
production asymmetries in hadron-nucleon collisions with several
different hadron beams at 250~GeV/c by the Fermilab E769
Collaboration \cite{E769}. The purpose of this work is to show
that the E769 data of $\Lambda$-$\bar{\Lambda}$ asymmetries in
hadron-nucleon collisions with different hadron beams may provide
useful information about the flavor structure of quark to
$\Lambda$ fragmentation functions.

We first look at the formalism for inclusive production of a
hadron $C$ from hadron $A$ and hadron $B$ collision process
\begin{equation}
A+B \to C+X,
\end{equation}
where the Mandelstam variables $s$, $t$, and $u$ are written as
\begin{eqnarray}
&s=(P_A+P_B)^2=M_A^2+M_B^2+ 2 P_A \cdot P_B,\\
&t=(P_A-P_C)^2=M_A^2+M_C^2-2 P_A \cdot P_C,\\
&u=(P_B-P_C)^2=M_B^2+M_C^2-2 P_B \cdot P_C.
\end{eqnarray}
The experimental cross sections are usually expressed in terms of
the experimental variables $x_F=2P_L/\sqrt{s}$ and $P_T$ at a
given $s$, where $P_L=x_F \sqrt{s}/2$ and $P_T$ are the
longitudinal and transversal momentum of the produced hadron $C$
with energy $E_C =\sqrt{M_C^2+P_C^2}=
\sqrt{M_C^2+P_L^2+P_T^2}=\sqrt{M_C^2+P_T^2+x_F^2 s/4}$ in the
center of mass frame of the collision process, and $s$ is the
squared center of mass energy.
In the collinear factorization theorem, the kinematics of the
subprocess $a +b \to c +d$ is defined as
\begin{eqnarray}
&p_a= x_a P_A,\\
&p_b= x_b P_B,\\
&p_c= P_C/ z,
\end{eqnarray}
from which we get the parton level Mandelstam variables $\hat{s}$, $\hat{t}$,
and $\hat{u}$
\begin{eqnarray}
&\hat{s}=(p_a+ p_b)^2=2 p_a \cdot p_b=x_a x_b s,\\
&\hat{t}=(p_a-p_c)^2=-2 p_a \cdot p_c=\frac{x_a}{z} t ,\\
&\hat{u}=(p_b-p_c)^2=-2 p_b \cdot p_c =\frac{x_b}{z} u .
\end{eqnarray}
The cross section can be written in terms of the parton level
subprocess as
\begin{equation}
\d \sigma=\frac{E_C \d ^3 \sigma^{AB \to C X}}{\d ^3 {\mathbf
P}_C} =\sum_{a,b,c,d}\int \frac{\d x_a \d x_b \d z }{\pi
z^2}f_{a}^{A}(x_a) f_{b}^{B}(x_b) \hat{s}
\delta(\hat{s}+\hat{t}+\hat{u})\frac{\d \hat{\sigma}^{a b \to c d
}}{\d \hat{t} }(x_a,x_b,z)D_{c}^{C}(z), \label{HardS}
\end{equation}
where $f_{a}^{A}(x_a)$ and $f_{b}^{B}(x_b)$ are the parton
distributions in the beam hadron $A$ and target proton $B$ with
Bjorken variables $x_a$ and $x_b$ respectively, $D_{c}^{C}(z)$ is
the fragmentation function of parton $c$ into the produced hadron
$C$ with energy fraction $z$ of hadron $C$ relative to the
scattered quark $c$, and $\frac{\d \hat{\sigma}^{a b \to c d }}{\d
\hat{t} }(x_a,x_b,z)$ is the subprocess cross section. From the
$\delta(\hat{s}+\hat{t}+\hat{u})$ function, we get the energy
fraction $z$
\begin{equation}
z=-\frac{x_a t+ x_b u}{x_a x_b s},
\end{equation}
and the two Bjorken variables for quarks $a$ and $b$ in hadrons
$A$ and $B$
\begin{equation}
x_a=\frac{-x_b u}{x_b z s+t}, \ \ x_b=\frac{-x_a t}{x_a z s+ u}.
\end{equation}
The integration over $x_a$ and $x_b$ can be done for both of $x_a$
and $x_b$ in the range $\left[0, 1\right]$, under the constraint
condition $0\le z \le 1$, or equivalently, for $x_a$ and $x_b$ in
the ranges $\left[x_a^{min},1\right]$ and
$\left[x_b^{min},1\right]$ respectively, with
\begin{equation}
x_{a}^{min}=-\frac{x_b u }{x_b s+t }, \ \ x_{b}^{min}=-\frac{t }{
s+u}.
\end{equation}

The hard scattering framework of Eq.~(\ref{HardS}) is perturbative
QCD based and therefore strictly valid only for high $P_T \geq
1$~GeV. Nevertheless, it has been recently argued that QCD
factorization is also applicable for semi-inclusive processes at
low transverse momentum \cite{JiMa}, therefore we make
extrapolations to smaller $P_T$ in this paper for the purpose of
illustration. The kinematical expressions that we use, however,
are different from the conventional expressions in terms of
rapidity \cite{CF} and are valid even for small $P_T$.

The cross sections of the subprocess at the parton level,
$\hat{\sigma}(\hat{s},\hat{t},\hat{u})$, can be found in
Refs.~\cite{BRST}. We adopt the Gl\"{u}ck, Reya, and Vogt (GRV)
leading order unpolarized parametrization for the nucleon parton
distributions~\cite{GRV95}. For the parton distributions of the
$K^+$ and $\pi^-$, we employ the parametrization forms of
Refs.~\cite{GRV-pion98} and \cite{PION99}, respectively. The above
quantities are well constrained by a vast number of available
experimental data, so we can focus our attention on the less known
quark to $\Lambda$ fragmentation functions.

We parametrize the quark to $\Lambda$ fragmentation functions
$D^{\Lambda}_q(z)$ by adopting the Gribov-Lipatov relation
\cite{GL}
\begin{equation}
D^{\Lambda}_q(z) \propto q^{\Lambda}(x), \label{GLR}
\end{equation}
in order to connect the fragmentation functions with the quark
distribution functions $q^{\Lambda}(x)$ of the $\Lambda$. More
explicitly, we adopt a general form to relate fragmentation and
distribution functions, as follows \cite{MSSY02PLB}
\begin{equation}
\begin{array}{ll}
D^{\Lambda}_V(z)=C_V(z) z^{\alpha}q^{\Lambda}_V(z),\\
D^{\Lambda}_S(z)=C_S(z) z^{\alpha}q^{\Lambda}_S(z),
\end{array}
\label{dfR}
\end{equation}
where a distinction between the valence ($V$) and the sea ($S$)
quarks is explicit. The above formulae are always correct, since
$C_V(z)$ and $C_S(z)$ are in principle arbitrary functions.  We
should consider Eq.~(\ref{dfR}) as a phenomenological
parametrization for the fragmentation functions of quarks and
antiquarks, as follows
\begin{equation}
\begin{array}{ll}
D^{\Lambda}_q(z)=D^{\Lambda}_V(z)+D^{\Lambda}_S(z),\\
D^{\Lambda}_{\bar{q}}(z)=D^{\Lambda}_S(z).
\end{array}
\label{Dqqbar}
\end{equation}
Three options were found \cite{MSSY02PLB} to fit quite well the
available experimental data of proton production in $e^+e^-$
inelastic annihilation: (1) $C_V=1$ and $C_S=0$ for $\alpha=0$;
(2) $C_V=C_S=1$ for $\alpha=0.5$; and (3) $C_V=1$ and $C_S=3$ for
$\alpha=1$. We adopt these three options to reflect the relation
between unfavored and favored fragmentation functions of the
$\Lambda$.

There is no direct measurement of the quark distributions of the
$\Lambda$. But we can relate the quark distributions between the
proton and the $\Lambda$ by assuming SU(3) symmetry between the
proton and the $\Lambda$ \cite{MSSY02SU3}
\begin{equation}
\begin{array}{lllc}
u_V^{\Lambda}(x)=d_V^{\Lambda}(x)=\frac{1}{6} u_V(x) + \frac{4}{6}
d_V(x),\\
s_V^{\Lambda}(x)=\frac{2}{3} u_V(x) - \frac{1}{3} d_V(x),
\end{array}
\end{equation}
for valence quarks, and
\begin{equation}
\begin{array}{lllc}
\bar{u}^{\Lambda}(x)=\bar{d}^{\Lambda}(x)=\frac{1}{2}
\left[\bar{u}(x)+\bar{s}(x)\right], \\
\bar{s}^{\Lambda}(x)=\bar{d}(x),
\end{array}
\end{equation}
for sea quarks. We adopt the GRV parametrization \cite{GRV95} of the
quark distributions $q(x)$ of the nucleon. In this way, we get a
complete set of quark distributions in the $\Lambda$ with both
valence and sea quark distributions.

It is well known that the flavor structure of $u$ and $d$ quark
distributions of the proton is different between the quark-diquark
model \cite{Fey72,DQM,Ma96} and a pQCD based analysis
\cite{Far75,Bro95}: the quark-diquark model predicts that
$d(x)/u(x) \to 0$ at $x \to 1$ whereas a pQCD based approach
predicts that $d(x)/u(x) \to 1/5$. A discrimination between the
two models requires very high precision measurement of the
structure functions at large $x$ and is difficult. On the other
hand, it has been also shown \cite{MSY00PLB} that this flavor
structure of the quark distributions at large $x$ is even more
significant in the case of the $\Lambda$,  with a large difference
between the ratio of $u^{\Lambda}(x)/s^{\Lambda}(x)$: the
quark-diquark model predicts that $u^{\Lambda}(x)/s^{\Lambda}(x)
\to 0$ at $x \to 1$, whereas the pQCD based approach predicts that
$u^{\Lambda}(x)/s^{\Lambda}(x) \to 1/2$. This will produce a large
difference in the ratio of fragmentation functions
$D_u^{\Lambda}(z)/D_{s}^{\Lambda}(z)$, which might be more easily
accessible experimentally via quark to $\Lambda$
fragmentation~\cite{ASY02}.

The valence quark distributions of the $\Lambda$ in the
quark-diquark model and the pQCD based analysis have been
explicitly studied \cite{MSY00PLB,MSY00Z,MSSY00} and we adopt the
parametrizations given in Ref.~\cite{MSSY00}. To describe the
$\bar{\Lambda}$ fragmentation, it is important to take into
account the sea contributions in the model construction. In order
to use the sea quark distributions from other parametrization
while still keeping the flavor structure of the valence quarks as
predicted in the two models, we re-scale the valence quark
distributions by a factor of
$u^{\Lambda}_{V,SU(3)}(x)/u^{\Lambda}_{V,th}(x)$, where the
subscript ``$SU(3)$" denotes the valence quark distributions of
the $\Lambda$ in the SU(3) symmetry model \cite{MSSY02SU3} and
``$th$" denotes the corresponding quantities predicted in the
quark-diquark model or the pQCD based analysis \cite{MSSY00}. This
is done in order to normalize the $\Lambda$ quark distributions to
well known proton quark distribution parametrizations. Notice that
the valence $u$-quark distribution then becomes that of the SU(3)
model, while the others get a rescaling factor. In this way we can
adopt the sea quark distributions from the SU(3) symmetry model as
the sea distributions in the quark-diquark model and the pQCD
based analysis, to reflect the contribution from the unfavored
fragmentation. Thus we get another two sets of quark to $\Lambda$
fragmentation functions, denoted as the quark-diquark model and
the pQCD based analysis later on. This procedure is done with the
main motivation of constructing realistic quark to $\Lambda$
phenomenological fragmentation functions, which has some features
coming from specific theoretical arguments, i.e., the
quark-diquark model and the pQCD based analysis.

With all of the above mentioned subprocess cross sections at the
parton level, parton distributions for both the beam and the
target, and also the quark to $\Lambda$ fragmentation functions,
we can calculate the $x_F$-dependent $\Lambda$-$\bar{\Lambda}$
asymmetries in hadron-nucleon collisions
\begin{equation}
A(x_F)=\int_{P_{T}^{min}}^{P_{T}^{max}} \d^2 \vec{P}_T
\left[\frac{\d^3 \sigma^{\Lambda}}{\d^3 p_{\Lambda}}-\frac{\d^3
\sigma^{\bar{\Lambda}}}{\d^3
p_{\bar{\Lambda}}}\right]/\int_{P_{T}^{min}}^{P_{T}^{max}} \d^2
\vec{P}_T \left[\frac{\d^3 \sigma^{\Lambda}}{\d^3
p_{\Lambda}}+\frac{\d^3 \sigma^{\bar{\Lambda}}}{\d^3
p_{\bar{\Lambda}}}\right],
\end{equation}
and those with $P_T^2$-dependence
\begin{equation}
A(P_T^2)=\int_{x_{F}^{min}}^{x_{F}^{max}} \d x_F \left[\frac{\d^3
\sigma^{\Lambda}}{\d^3 p_{\Lambda}}-\frac{\d^3
\sigma^{\bar{\Lambda}}}{\d^3
p_{\bar{\Lambda}}}\right]/\int_{x_{F}^{min}}^{x_{F}^{max}} \d x_F
\left[\frac{\d^3 \sigma^{\Lambda}}{\d^3 p_{\Lambda}}+\frac{\d^3
\sigma^{\bar{\Lambda}}}{\d^3 p_{\bar{\Lambda}}}\right].
\end{equation}
We do not introduce any additional parameters to fit the data for
the input quantities.

The E769 Collaboration measured the asymmetries with 250~GeV/c
$\pi^{\pm}$, $K^{\pm}$, and $p$ beams on the proton target. The
data are expressed as functions of $x_F$  and $P^2_T$ over the
ranges $-0.12\le x_F \le 0.12$ and $0 \le P^2_T\le  3
~(\mbox{GeV/c})^2$ for positively charged beam, and $-0.12\le x_F
\le 0.4$ and $0 \le P^2_T\le  10 ~(\mbox{GeV/c})^2$ for negatively
charged beam. In fact, the $x_F$-dependence was measured without
$P_T^2$ cut, and the $P_T^2$-dependence was measured without $x_F$
cut. Notice that in the $P_T$ integration we have considered the
small ($\leq 1$ GeV) region, where the hard scattering formalism
is not strictly valid. We have checked that the final results are
not really sensitive to this lower limit.

\begin{figure}
\begin{center}
\leavevmode {\epsfysize=16cm \epsffile{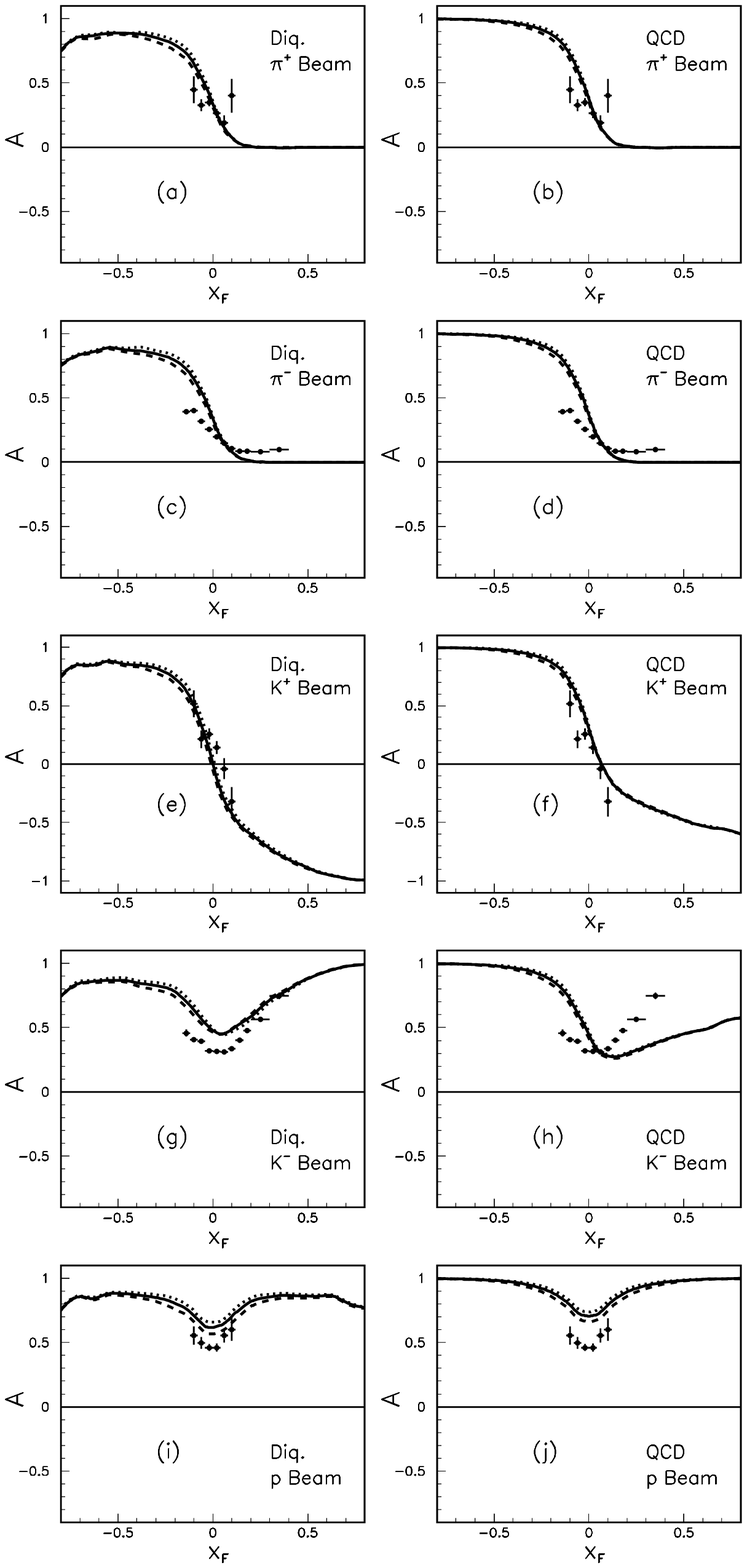}}
\end{center}
\caption[*]{\baselineskip 13pt The prediction for the $\Lambda$
production asymmetries vs $x_F$ for various incident beams, and
with three options for the contribution from unfavored
fragmentation: (1) the dotted curve is for $C_V=1$ and $C_S=0$,
$\alpha=0$;(2) the solid curve is for $C_V=C_S=1$, $\alpha=0.5$;
and (3) the dashed curve is for $C_V=1$ and $C_S=3$, $\alpha=1$.
The left row figures correspond to results from the quark-diquark
model, and the right row figures correspond to results from a pQCD
based analysis. The integration range of $P_T^2$ is $0 \le
P^2_T\le 3 ~(\mbox{GeV/c})^2$ for positively charged beams, and $0
\le P^2_T\le  10 ~(\mbox{GeV/c})^2$ for negatively charged beams.
The experimental data are taken from Ref.~\cite{E769}.  }
\label{msy16f1}
\end{figure}

\begin{figure}
\begin{center}
\leavevmode {\epsfysize=16cm \epsffile{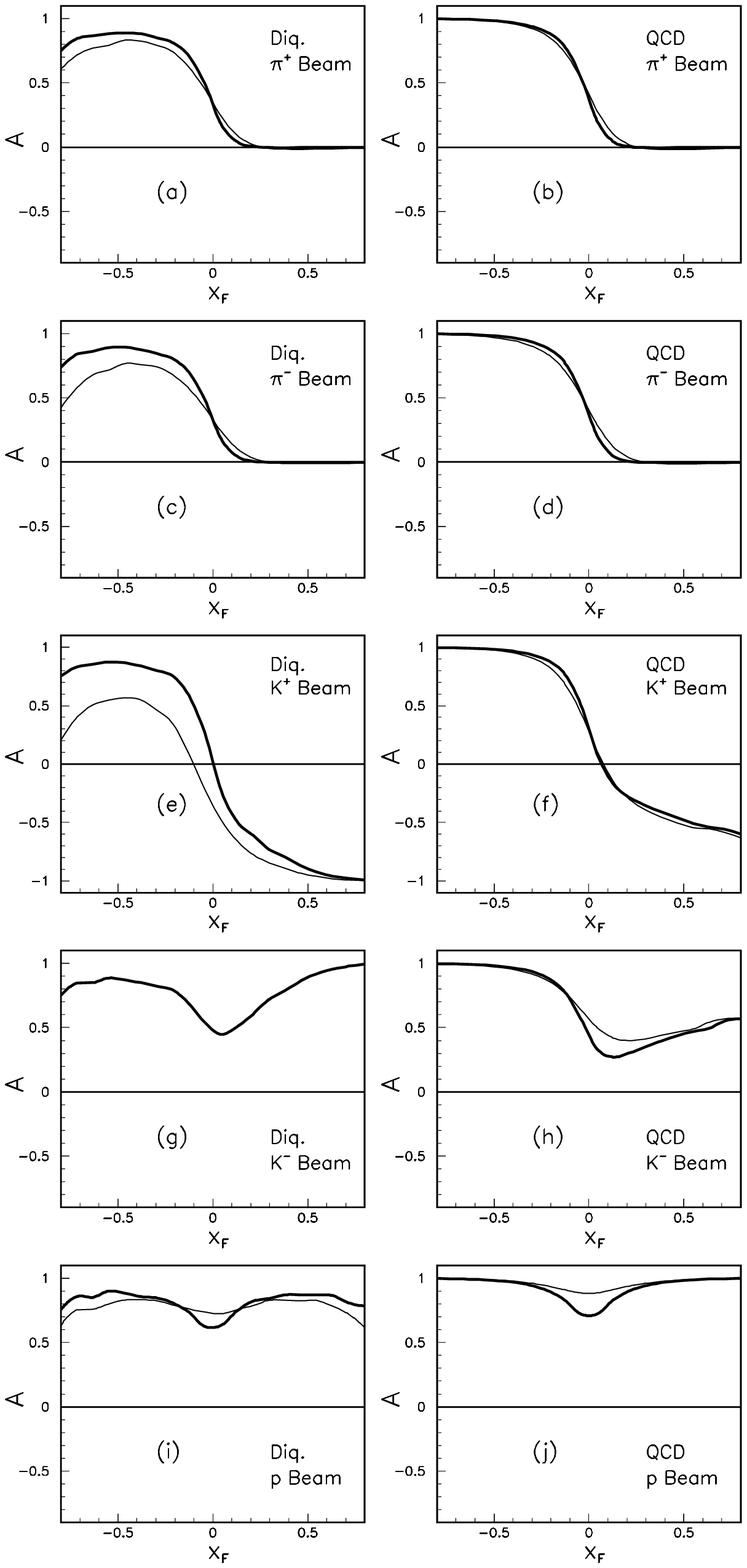}}
\end{center}
\caption[*]{\baselineskip 13pt The same as in Fig.~\ref{msy16f1},
using only the contribution from unfavored fragmentation with
$C_V=C_S=1$ for $\alpha=0.5$. The integration range of $p_T^2$ is
[0, 5] GeV$^2$ for the thick solid curves and [5, 10] GeV$^2$ for
the thin solid curves. } \label{msy16f2}
\end{figure}

In the theoretical calculation, we need to set the limits of
$P_T^{min}$ and $P_T^{max}$ for the $x_F$-dependence, and of
$x_F^{min}$ and $x_F^{max}$ for the $P_T^2$-dependence. We will
choose the above mentioned ranges in the calculation. In
Fig.~\ref{msy16f1}, we plot the calculated results for the
$x_F$-dependent asymmetries $A(x_F)$, and find that both the
quark-diquark model and the pQCD based analysis can explain the
trend of the data. The quark-diquark model seems to be somewhat
more favored by the data for $K^{\pm}$ beams, especially at large
$x_F$. The differences between the quark-diquark model and the
pQCD based analysis can be understood by the fact that $x_F
\approx x_a-x_b$ in the Bjorken limit, and that the large $x_F$
behaviors is mainly controlled by the $\bar{s}$ and $s$ quark
fragmentation of the beam at large $x_F$. In the quark-diquark
model, the $s$ quark fragmentation dominates over the light-flavor
quark fragmentation. Therefore the dominant $\bar{s}$
fragmentation of the $K^+$ beam favors $\bar{\Lambda}$ production,
whereas the dominant ${s}$ fragmentation of the $K^-$ beam favors
${\Lambda}$ production. This can explain why the asymmetries
$A(x_F)$ for the $K^-$ beam is negative at large $x_F$ whereas
they are positive for the $K^+$ beam. From the different flavor
structure of $D^{\Lambda}_u(z)/D^{\Lambda}_s(z) \to 0$ in the
quark-diquark model and of $D^{\Lambda}_u(z)/D^{\Lambda}_s(z) \to
1/2$ in the pQCD based analysis, we can explain why the
quark-diquark model predicts large magnitude for the asymmetries
$A(x_F)$ at large $x_F$ for $K^{\pm}$ beams, since the $s$ quark
fragmentation dominates over the light-flavor fragmentation in the
quark-diquark model. We thus conclude that the large $x_F$
behavior of the $\Lambda$-$\bar{\Lambda}$ asymmetries can
discriminate between the two model predictions. We present in
Fig.~\ref{msy16f2} of the calculated results of the $A(x_F)$
asymmetries with two different options of $P_T$-cut: the
integrated ranges of $[(P_T^{\min})^2,(P_T^{\max})^2]$ are $[0,5]$
and $[5,10]$ respectively. We find that the trend in the
$x_F$-dependence does not change much for the two options,
although the magnitudes are different.

\begin{figure}
\begin{center}
\leavevmode {\epsfysize=16cm \epsffile{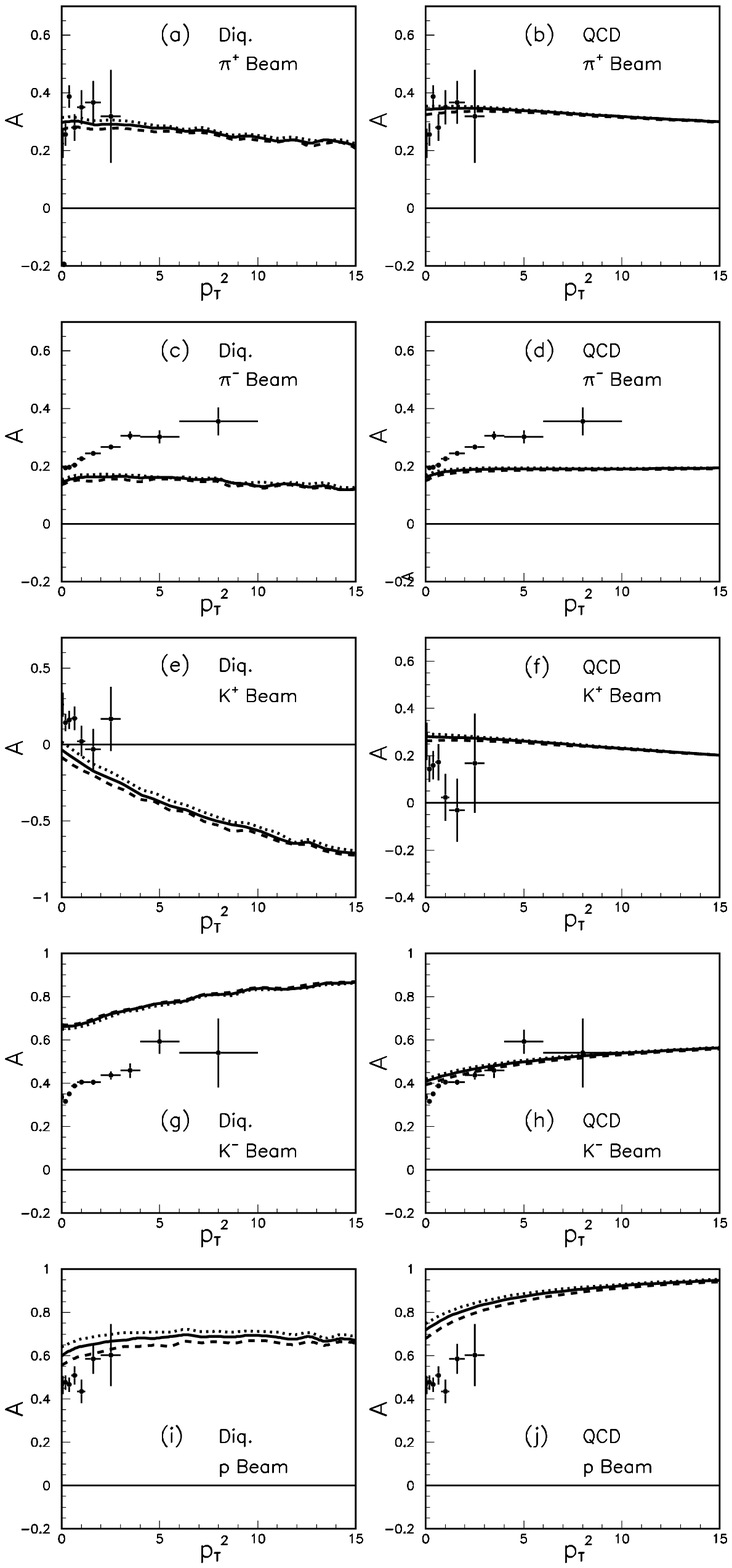}}
\end{center}
\caption[*]{\baselineskip 13pt The prediction of
 $\Lambda$ production asymmetries
vs $P_T^2$ for various incident beams with the dotted, solid, and
dashed curves corresponding to the three options for the unfavored
fragmentation as in Fig.~\ref{msy16f1}. The integration range of
$x_F$ is $-0.12\le x_F \le 0.12$ for positively charged beams, and
$-0.12\le x_F \le 0.4$ for negatively charged beams. The
experimental data are taken from Ref.~\cite{E769}. }
\label{msy16f3}
\end{figure}

\begin{figure}
\begin{center}
\leavevmode {\epsfysize=16cm \epsffile{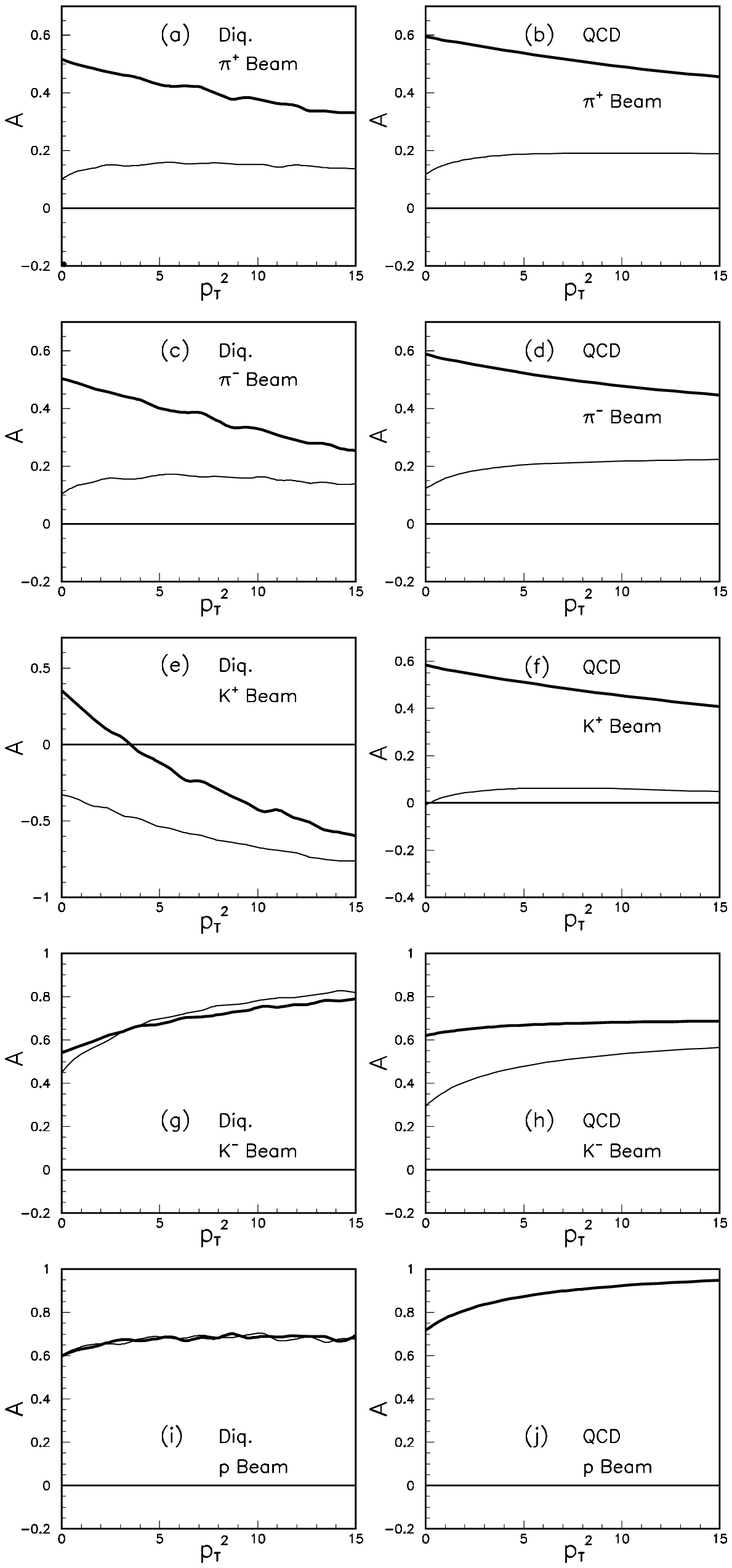}}
\end{center}
\caption[*]{\baselineskip 13pt
The same as in Fig.~\ref{msy16f3}
by using only the contribution from unfavored fragmentation
with $C_V=C_S=1$ for $\alpha=0.5$. The integration
range of $x_F$ is [-0.12, 0.0]  for the thick solid curves and
[0.0, 0.12] for the thin solid curves.
} \label{msy16f4}
\end{figure}

We also plot the calculated results for the $P_T^2$-dependent
asymmetries $A(P_T^2)$ in Fig.~\ref{msy16f3}, and find that both
the quark-diquark model and the pQCD based analysis can explain
the $P_T^2$-dependence trend, but the results differ slightly in
magnitude from the data. The reason is that the asymmetries
$A(P_T^2)$ are very sensitive to the $x_F$ integration range
$[x_F^{\min},x_F^{\max}]$, and we cannot reproduce exactly the
real $[x_F^{\min},x_F^{\max}]$ of the data. This is supported by
Fig.~\ref{msy16f4}, in which we plot the calculated results of
$A(P_T^2)$ with two options of $x_F$-cut: the integrated ranges of
$[x_F^{\min},x_F^{\max}]$ are $[-0.12,0]$ and $[0,0.12]$
respectively. We find that the magnitude of the asymmetries
$A(P_T^2)$ differ significantly for the two options, whereas the
trend of the $P_T^2$-dependence is similar. Thus both the
quark-diquark model and the pQCD based analysis can explain the
$P_T^2$-dependence in trend, but the exact magnitude is sensitive
to the $x_F$-cut. Therefore we conclude that the magnitude of
$A(P_T^2)$ with detailed $x_F$-cut information is important for
the purpose of distinguishing between different model predictions.
Thus far we only expect our formalism to work for asymmetries,
rather than for cross sections. In this last case there are
questions of normalization and evolution which are not present for
asymmetries.

In summary, we showed in this work that the
$\Lambda$-$\bar{\Lambda}$ asymmetries in hadron-nucleon collisions
can provide information about the flavor structure of quark to
$\Lambda$ fragmentation functions. Both the quark-diquark model
and a pQCD based analysis can explain the available data
qualitatively. We also showed that the two models give significant
different predictions for the magnitude of the $x_F$-dependent
asymmetries at large $x_F$ for $K^{\pm}$ beams. Thus high
precision measurements of the asymmetries with detailed $x_F$-cut
and $P_T$-cut information are important to  discriminate between
different predictions.

{\bf Acknowledgments}

We acknowledge the helpful communication with H.~da Motta for some
detailed information of the E769 data. This work is partially
supported by National Natural Science Foundation of China under
Grant Numbers 10025523 and 90103007, by Fondecyt (Chile) grant
1030355, by Alexander von Humboldt-Stiftung (J.-J.~Yang), and by
Foundation for University Key Teacher by the Ministry of Education
(China).

\end{document}